# Characterization of $Er^{3+}$:$YVO_4$ for microwave to optical transduction

Tian Xie[1,2], Jake Rochman[1,2], John G. Bartholomew[1,2,3],
Andrei Ruskuc[1,2], Jonathan M. Kindem[1,2,4], Ioana Craiciu[1,2,5], Charles Thiel[6], Rufus Cone[6],
Andrei Faraon[1,2,*]

[1]Kavli Nanoscience Institute and Thomas J. Watson, Sr., Laboratory of Applied Physics,
California Institute of Technology, Pasadena, California 91125, USA
[2]Institute for Quantum Information and Matter, California Institute of Technology, Pasadena,
California 91125, USA
[3]Current address: Centre of Engineered Quantum Systems, School of Physics, The University of
Sydney, Sydney, NSW 2006, Australia
The University of Sydney Nano Institute, The University of Sydney, NSW 2006, Australia
[4]Current address: JILA, University of Colorado and NIST, Boulder, CO, USA; Department of
Physics, University of Colorado, Boulder, CO, USA; National Institute of Standards and
Technology (NIST), Boulder, CO, USA
[5]Current address: Jet Propulsion Laboratory, California Institute of Technology, Pasadena,
California 91125, USA
[6]Department of Physics, Montana State University, Bozeman, Montana 59717, USA

Quantum transduction between microwave and optical frequencies is important for connecting superconducting quantum platforms in a quantum network. Ensembles of rare-earth ions are promising candidates to achieve this conversion due to their collective coherent properties at microwave and optical frequencies. Erbium ions are of particular interest because of their telecom wavelength optical transitions that are compatible with fiber communication networks. Here, we report the optical and electron spin properties of erbium doped yttrium orthovanadate ($Er^{3+}$:$YVO_4$), including high-resolution optical spectroscopy, electron paramagnetic resonance studies and an initial demonstration of microwave to optical conversion of classical fields. The highly absorptive optical transitions and narrow ensemble linewidths make $Er^{3+}$:$YVO_4$ promising for magneto-optic quantum transduction.

# I: Introduction

Advances in quantum technologies are emerging quickly, including demonstrations of quantum computation and simulation using superconducting qubits and atomic qubits [1,2], distribution of entangled optical photons over long distances [3,4] and quantum memories based on solid state or atomic ensembles [5,6]. Incorporating the best technologies from different physical systems into a single network requires coherent transfer of quantum information between different operating regimes. One way to achieve this is to build a quantum transducer. A primary example is to convert quantum states encoded in microwave photons to optical frequencies, which would enable distributed quantum computing schemes based on superconducting qubits or spin qubits [7]. Many physical systems have been proposed for microwave to optical (M2O) transduction [8,9], including optomechanical systems [10,11], electro-optical systems [12,13], atomic ensembles [14,15] and others [16,17].

Among the atomic ensemble approaches, rare-earth ions (REIs) in solids are a promising platform for M2O transduction applications [14,18,19,20,21]. REIs can be optically addressed using their narrow 4f-4f transitions, while electron spin, nuclear spin, or magnonic transitions can be used in the microwave domain. Although the optical absorption of a single REI is relatively weak, ensembles of REIs can exhibit large optical depths due to the narrow inhomogeneities at cryogenic temperatures [22,23]. Additionally, REIs doped in crystals possess long coherence lifetimes in both the optical and spin domain [24,25], which gives the possibility of a built-in memory incorporated with the transducer. Lastly, isotopes of REIs with nonzero nuclear spin have zero-field hyperfine structure [20,26], which enables transduction without an external magnetic field.

One promising transduction protocol involves using a cavity-enhanced Raman scattering process with a 3-level system [18]. In this scheme, high transduction efficiency can be achieved when the product of the optical and spin cooperativities is large, which requires ensembles possessing large transition strengths and narrow inhomogeneities in both optical and microwave domain. More specifically, in the limit of adiabatic driving and low efficiency ( < $10^{-2}$), the transduction efficiency scales as $\zeta \equiv \left(\frac{d_{31}d_{32}\mu_{21}\rho}{\Delta_o \Delta_\mu}\right)^2$ [17], where $d_{ij}$ ($\mu_{ij}$) is the optical (spin) dipole moment between levels $i$ and j of the 3-level system, $\rho$ is the number density of REIs, and $\Delta_{o(\mu)}$ is the detuning from the atomic resonance in the optical (microwave) domain, as shown in Fig. 1. This relationship indicates there is a strong dependence of the transduction efficiency on the specific REI and host crystal.

Among the REIs, erbium is an attractive element because its $^4I_{15/2}$ - $^4I_{13/2}$ optical transitions occur in the lowest-loss telecommunication band for fiber-based optical communication networks. Erbium has been extensively studied for quantum information technologies in host crystals including $Y_2SiO_5$ [27,28,29] and there has been recent work in other hosts such as $YPO_4$ [30], $Y_2O_3$ [31], $LiYF_4$ [32], $LiNbO_3$[33] and $TiO_2$[34]. Yttrium orthovanadate ($YVO_4$) is an attractive host crystal because of its high site symmetry which leads to first order insensitivity of optical transition frequencies to electric field and strain, that results in narrow inhomogeneous lines at relatively high concentrations [23]. In this work, we investigate erbium-doped $YVO_4$ ($Er^{3+}$:$YVO_4$) for its potential application in quantum transduction. Previous studies of $Er^{3+}$:$YVO_4$ have shown crystal-field analysis [35], lasing properties [36] and absorption and relaxation dynamics [37,38] down to 4 K with low frequency resolution (>100 GHz - equivalent to nanometer-level). Recently, $^{167}Er^{3+}$:$YVO_4$ coherent dynamics were investigated and showed a 15s ground state hyperfine lifetime, indicating that $^{167}$Er$^{3+}$:YVO$_4$ is a promising material for quantum memory applications [39].

In this work, we present the optical and spin properties of even (zero nuclear spin) isotope Er$^{3+}$:YVO$_4$ at 1 K. We characterize the optical transition strength, optical inhomogeneity and the electronic g tensors using high-resolution (~1 MHz) optical spectroscopy and measure the ground state spin inhomogeneity using electron paramagnetic resonance (EPR) techniques. The highly absorptive optical transitions and narrow inhomogeneities make this material a promising material for REI quantum transducers, which we explore further through coherent M2O conversion of classical signals.

This paper is organized as follows: Section II provides details on the material properties, energy level structure, selection rules and the spin Hamiltonian studied in this work. Section III introduces the experimental setup. Section IV presents the experimental results including optical transmission spectroscopy, EPR measurements and the M2O transduction measurements.

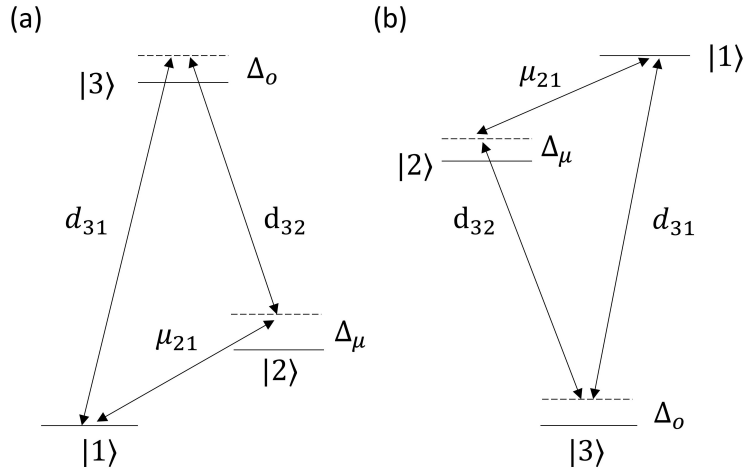

Fig. 1: Energy level diagram of a 3-level system for the cavity enhanced transduction process. (a) a $\Lambda$ system (b) a $V$ system. $d_{ij}$ ($\mu_{ij}$) is the optical (spin) dipole moment between level $i$ and $j$, $\Delta_{o(\mu)}$ is the detuning from the atomic resonance in optical (microwave) domain.

## **II: Background**

YVO$_4$ is a zircon tetragonal crystal with D$_{4h}$ symmetry. The unit-cell parameters are a = b = 7.1183 Å and c = 6.2893 Å [40]. There is one yttrium site in the unit cell with local D$_{2d}$ point group symmetry, which can be substituted by an erbium ion. Due to the non-polar symmetry at the substitutional site, there is no first-order DC Stark effect, which results in lower sensitivity to electric field noise that can cause decoherence and spectral diffusion of optical transitions [20]. The uniaxial site symmetry greatly reduces the parameters needed to characterize the system compared to some alternative host crystals with lower symmetry (for example in Y$_2$SiO$_5$).

Er$^{3+}$ has 11 electrons in the 4f shell, which gives the first two spin-orbit multiplets as $^4$I$_{15/2}$ and $^4$I$_{13/2}$, from the free ion Hamiltonian [41]. The crystal field with D$_{2d}$ symmetry splits the energy levels of the free Er$^{3+}$ ion, which are characterized by the total angular momentum J, into J+1/2 Kramers doublets. The $^4$I$_{15/2}$ ($^4$I$_{13/2}$) level splits into 8 (7) levels, labelled as Z$_1$ to Z$_8$ (Y$_1$ to Y$_7$) in order from the lowest to the highest energy, as showed in Fig. 2(a). These labels transform to the irreducible representations $\Gamma_6$ or $\Gamma_7$ and can be described by the crystal field quantum number $\mu = \pm\frac{1}{2}, \pm\frac{3}{2}$ according to the $D_{2d}$ point group [42,43]. From the irreducible representations, the selection rules can be determined [32] as shown in Fig. 2(b) and (c), where the $\pi$ ($\sigma$) represents the polarization $\mathbf{E}_{ac}\|c$ ($\mathbf{E}_{ac}\|a$) and ED (MD) denotes the electric (magnetic) dipole transition. For a small applied magnetic field $\mathbf{B}$, the Kramers doublets can be described using an effective spin Hamiltonian with $S = 1/2$ [41]:

$$\mathcal{H}_{eff} = \mu_B \mathbf{B} \cdot \mathbf{g} \cdot \mathbf{S}$$

where $\mathbf{g}$ is the electronic Zeeman tensor and $\mathbf{S}$ is the effective spin-1/2 operator. Because of the local $D_{2d}$ symmetry, only the diagonal terms of the $\mathbf{g}$ tensor are non-zero, which we label $g_\|$ and $g_\perp$ for components parallel and perpendicular to the crystal symmetry c-axis, respectively.

At cryogenic temperature, only the $Z_1$ level is populated because the crystal field splitting in the ground state is approximately 1.1 THz ($\hbar\omega/k_b \sim 50$ K $\gg$ 1 K). The splitting between $Y_1$ and $Y_2$ is only 55 GHz at zero-field, which may be detrimental for coherent optical properties of $Y_1$ and $Y_2$ at high temperature (eg: > 4 K) due to phonon relaxation processes [44] but should be less of a problem for quantum transduction experiments that operate at T<100 mK. The following experiments and discussions are focused on the levels $Z_1$, $Y_1$ and $Y_2$. The optical transitions between these levels are around 1530 nm.

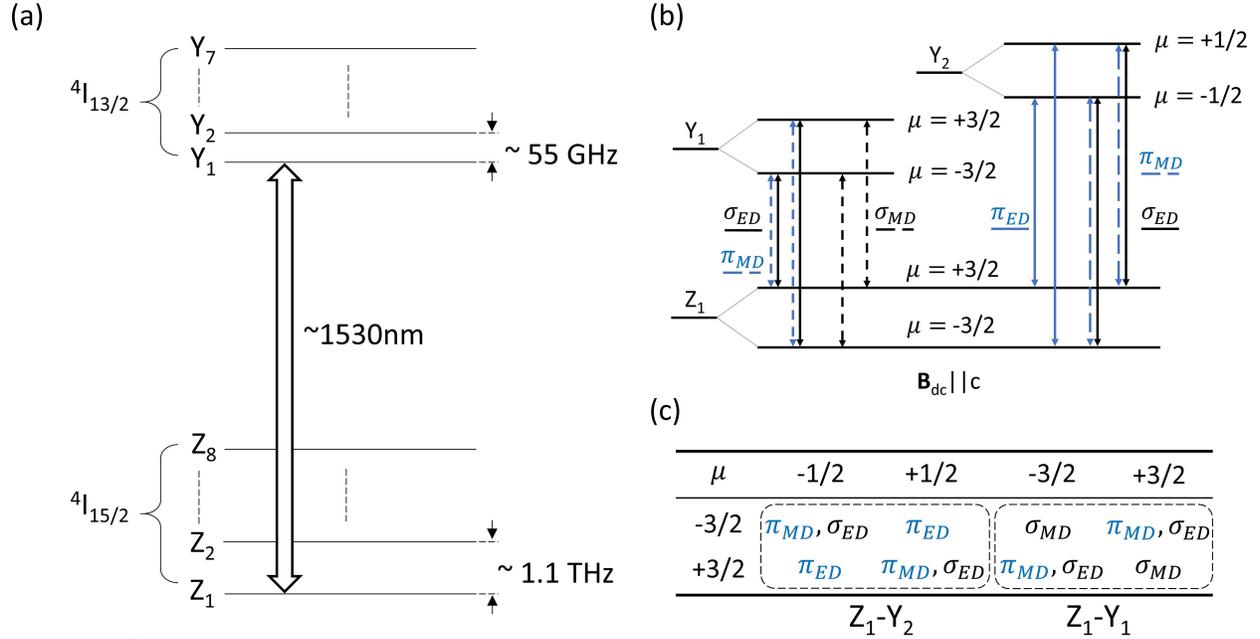

Fig. 2: $Er^{3+}$:$YVO_4$ energy levels and selection rules. (a) The energy-level diagram of $Er^{3+}$:$YVO_4$ at zero applied magnetic field. (b) Partial diagram of $Er^{3+}$:$YVO_4$ crystal field and Zeeman levels with the crystal field quantum number and the selection rules under external magnetic field along the crystal symmetry c-axis ($\mathbf{B}_{dc}\|c$). $\pi(\sigma)$ indicates the polarization $\mathbf{E}_{ac}\|c$ ($\mathbf{E}_{ac}\|a$) with blue (dashed) line styles. ED (MD) denotes the electric (magnetic) dipole transition. (c) The selection rules for electric and magnetic dipole moment transitions in $D_{2d}$ symmetry.

## III: Experimental Setup

Samples were cut and polished from a boule of $YVO_4$ doped with a natural abundance of $Er^{3+}$ grown by Gamdan Optics, which is comprised of 77% nuclear-spin-0 even isotopes and 23% odd isotope $^{167}Er^{3+}$ that has nuclear spin 7/2. The even isotope $Er^{3+}$ concentration was 140 ppm (measured by Secondary Ion Mass Spectrometry). For the main experiments, we used a 200 μm thick a-cut crystal mounted within a 2.4 GHz loop-gap microwave resonator (Q = 860) machined from oxygen-free copper sitting on the still plate (base temperature of 1 K) of a dilution refrigerator, as shown in Fig. 3. The loop-gap resonator sat on a fiber coupled U-bench (Thorlabs FBC-1550-APC) with two fiber collimators (Thorlabs PAF-X-2-C) for optical transmission measurements. The total optical coupling efficiency through the U-bench setup was 35% at 1 K, limited by misalignment due to thermal contraction during cooldown. The light propagation direction was along the a-axis of the crystal, which allowed measurement of both $\pi$ and $\sigma$ polarized spectra. A DC magnetic field ($\mathbf{B}_{dc}$) was applied to the crystal along the c-axis using a home-built split coil superconducting magnet that generated fields up to 120 mT. To determine the electronic Zeeman $g$ tensors for levels $Z_1$, $Y_1$ and $Y_2$, we placed the crystal inside a 2-axis superconducting magnet to perform magnetic field rotations with respect to the crystal symmetry axes (not shown in Fig.4 schematic).

Optical signals, as shown using solid lines in the lower part of the circuit diagram in Fig. 3, were generated from a tunable diode laser (Toptica CTL). The absolute laser frequency was calibrated using a wavemeter (Bristol 671A). The laser light was split into two paths. The signal path, containing 10% of the power, went through a polarization controller (POL) and a fiber acousto-optic modulator (AOM), to perform pulsed measurements. The 90% path was used as the local oscillator for heterodyne measurements. For continuous wave optical measurements, the light transmitted through the crystal was measured with a photodiode. For pulsed lifetime measurements, another fiber AOM was used to shutter a superconducting nanowire single-photon detector (SNSPD) which detected the fluorescence signal. Microwave signals, indicated by dashed lines in the upper part of the circuit diagram in Fig. 3, were coupled into and out of the loop-gap resonator using two electric dipole antennas positioned within the loops [45]. For electron paramagnetic resonance (EPR) measurements, we used a function generator (TPI-1002-A) to generate a 100 kHz frequency-modulated (FM) microwave signal centered at the loop-gap resonator frequency (2.4 GHz). A RF power meter (RF Bay RPD-5513) was used to detect and mix down the FM signal that was transmitted through the antennas. The beat-note amplitude was measured using a spectrum analyzer (FieldFox N9915A). For the M2O transduction experiment, a microwave tone was generated from the spectrum analyzer and a laser tone was sent to the crystal. For detection, we combined the collected light with the local oscillator path and detected the interference on a fast photodiode (UPD-35-IR2-FC). The beat-note signal was amplified (WBA2080-35A) and recorded using the spectrum analyzer.

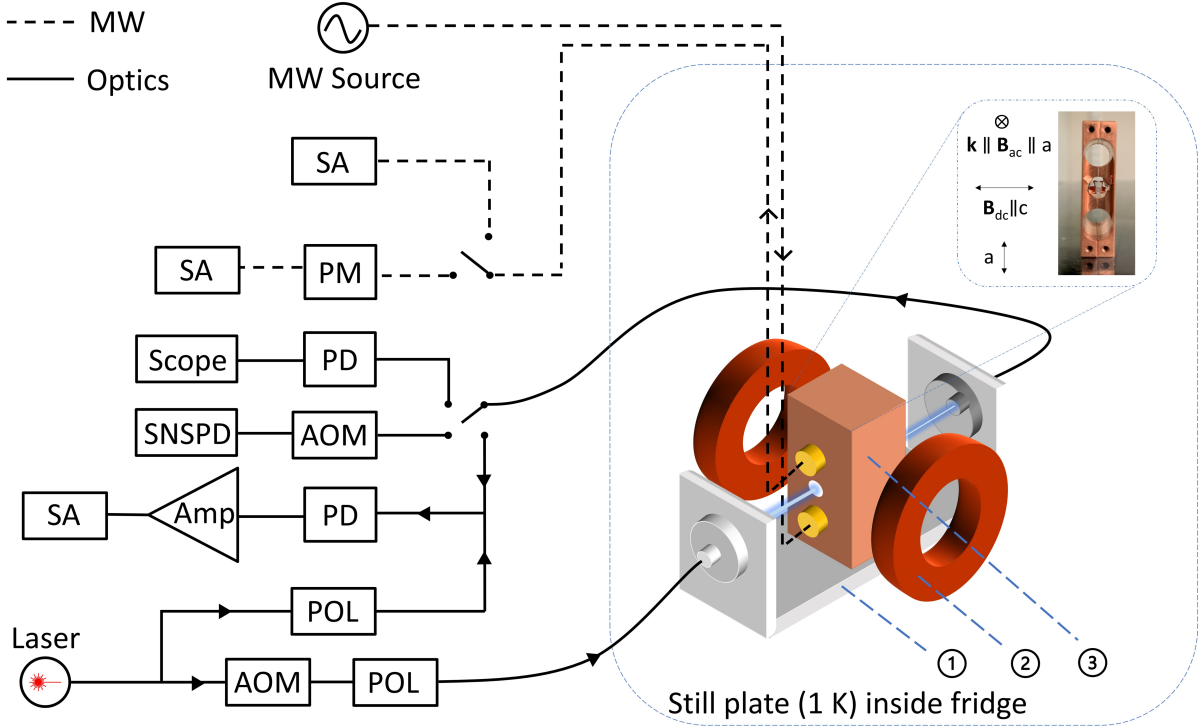

Fig. 3: Experimental setup inside and outside the dilution fridge for optical and microwave measurements. ①: fiber-to-fiber U-bench, ②: superconducting magnet, ③: loop-gap resonator, AOM: acousto-optic modulator, POL: polarization controller, PD: photodiode, SNSPD: superconducting nanowire single-photon detector, Amp: amplifier, SA: spectrum analyzer, PM: power meter, MW: microwave. Solid lines show the optical path, and dashed lines indicate electronic connections. The light propagates along the crystal a-axis, which allows $\pi$ and $\sigma$ excitation. $\mathbf{B}_{dc}$ is along the crystal c-axis. The direction of the oscillating magnetic field ($\mathbf{B}_{ac}$) generated by the loop-gap resonator is along the crystal a-axis.

## IV: Experimental Results
### A. Optical transmission spectrum

High-resolution optical transmission spectra for $Z_1$-$Y_1$ and $Z_1$-$Y_2$ were performed as a function of the bias magnetic field for $\mathbf{B}_{dc}\|c$ (see Fig. 4(a)-(d)). We observed narrow inhomogeneously broadened and highly absorptive transition lines. For $Z_1$-$Y_1$ under $\sigma$ polarized excitation, there are four well-resolved transitions with an average inhomogeneous linewidth of $184\pm10$ MHz, as shown in Fig. 4(a). The number of allowed transitions is consistent with the theoretical calculation, where four (two) lines are observed under $\sigma$ ($\pi$) polarization (see Fig. 2(b)). For $Z_1$-$Y_2$ with $\pi$ polarized excitation (Fig. 4(b)), we observed four highly absorbing transitions with an average inhomogeneous linewidth of $163\pm14$ MHz. The ratio of the absorption strength from same dipole moments connecting different ground-state spin levels gives the thermal distribution between ground-state spin levels, which was consistent with a 1 K system temperature. Additional absorption lines were observed from the $^{167}$Er isotope (40 ppm). These resolved optical transitions allow us to individually address transitions between different Zeeman levels. Also, the polarizations with all four allowed transitions provide 3-level systems for transduction [18] as described in the introduction section.

The total integrated absorption coefficient for $Z_1$-$Y_1$ ($Z_1$-$Y_2$) is 97.2 (125) GHz·cm$^{-1}$ for $\sigma$ ($\pi$) polarized excitation, from which the electric dipole (ED) and magnetic dipole (MD) absorption oscillator strength, $f_{ij,q}$, of the transitions can be calculated using [23,25]:

$$f_{ij,q}^{ED} = 4\pi\epsilon_0 \frac{m_e c}{\pi e^2} \frac{1}{N} \frac{n_q}{\chi_L} \int \alpha_q^{ED}(\nu) d\nu, \qquad f_{ij,q}^{MD} = 4\pi\epsilon_0 \frac{m_e c}{\pi e^2} \frac{1}{N} \frac{1}{n_q} \int \alpha_q^{MD}(\nu) d\nu$$

where $\epsilon_0$ is the vacuum permittivity, $m_e$ is the electron mass, $c$ is the speed of the light, $e$ is the charge of an electron, $N$ is the number density, $q$ is the polarization, $n_q$ is the refractive index along different directions, $\chi_L = \left(\frac{n_q^2+2}{3}\right)^2$ is the local electric field correction factor (virtual cavity model), and $\alpha_q$ is the absorption coefficient for different polarizations. Because there is no degeneracy of the Zeeman levels here, the emission oscillator strength is the same as the absorption oscillator strength $f_{ji} = f_{ij}$. For YVO$_4$, the refractive index along the c (a) axis is 2.15 (1.95) [46]. With the doping concentration equal to 140 ppm, the even-isotope erbium dopant number density is $N = 1.75 \times 10^{18}$ cm$^{-3}$, distributed between the two Zeeman levels. The corresponding dipole moment was calculated from the oscillator strength using $d_{ij,q}^2 = \frac{\hbar e^2}{2m\omega} f_{ij,q}$. The radiative lifetime was calculated using [23,25]:

$$\frac{1}{\tau_{rad,q}^{ED}} = \frac{2\pi e^2}{\epsilon_0 m_e c} \frac{\chi_L n_q}{\lambda_0^2} \frac{f_{ji,q}^{ED}}{3}, \qquad \frac{1}{\tau_{rad,q}^{MD}} = \frac{2\pi e^2}{\epsilon_0 m_e c} \frac{n_q^3}{\lambda_0^2} \frac{f_{ji,q}^{MD}}{3}$$

where $\lambda_0$ is the transition wavelength in vacuum. The calculated absorption coefficient integral, oscillator strength, radiative lifetime, and dipole moment for different transitions are listed in Table 1. From these numbers, we obtain the total dipole moment ($d_{ij}^2 = \sum_q d_{ij,q}^2$) as $d_{ED} = 1.4(3.5) \times 10^{-32}$ C·m and $d_{MD} = 4(3.5) \times 10^{-32}$ C·m for the $Z_1$-$Y_1$ ($Z_1$-$Y_2$) transition. The measured MD transition oscillator strength agrees with the theoretical calculation in [47] within a factor of 2. From the calculated radiative lifetimes shown in Table 1, we obtain the total radiative rate $\frac{1}{\tau_{rad}} = \sum_q \left(\frac{1}{\tau_{rad,q}^{ED}} + \frac{1}{\tau_{rad,q}^{MD}}\right)$ to be 1/8.1ms for $Y_1$-$Z_1$ and 1/6.2ms for $Y_2$-$Z_1$ transitions.

To determine the electronic Zeeman $g$ factors, we measured the $Z_1$-$Y_1$ and $Z_1$-$Y_2$ optical transition frequencies as a function of the bias magnetic field $\mathbf{B}_{dc}$, shown in Fig. 4(c) and (d). The red dashed lines are derived from an effective spin Hamiltonian fitted to the data. With the knowledge of the electronic Zeeman $g$ factors ($g_\parallel$ = 3.544, $g_\perp$ = 7.085) for $Z_1$ [48], we obtain the excited state $g$ factors: $g_\parallel$ = 4.51 ± 0.02 for $Y_1$ and 2.74 ± 0.01 for $Y_2$. In order to determine $g_\perp$, we performed a magnetic field rotation experiment at a field strength around 75 mT, shown in Fig. 4(e) and (f), where the field direction is rotated from the c-axis to the a-axis. By fitting the data to an effective spin Hamiltonian model, we get $g_\perp$ = 4.57±0.01 for $Y_1$ and 6.74±0.15 for $Y_2$. The $g$ factors are summarized in Table 2. We attribute the deviation of the fitting results near 90 degrees in the $Z_1$-$Y_2$ rotation pattern to the quadratic Zeeman effect [49], which is not included in the model.

To investigate $Y_i$-$Z_1$ branching ratios, we measured the fluorescence decay lifetimes when exciting ensembles to $Y_1$ or $Y_2$ at 1 K. To minimize the impact of radiation trapping [50], which is present at the center of the inhomogeneous line, we excited the ensemble with the laser frequency detuned by two inhomogeneous linewidths from the center of the relevant absorption peak. The fluorescence data detected on the SNSPD is shown in Fig. 5. When population was excited to $Y_1$ ($Y_2$), the extracted lifetime $\tau_f$ was 3.34 ± 0.01 ms (3.30 ± 0.01 ms). Given the 55 GHz energy separation between $Y_1$ and $Y_2$, the phonon relaxation between the states will significantly modify the branching ratios to $Z_1$ as the temperature is varied. In the low temperature limit ($kT \ll \hbar\Delta$) phonon absorption from $Y_1$ to $Y_2$ is suppressed and the branching ratio for the $Y_1$-$Z_1$ transition will be at least $\beta_{Y1} = \tau_{f\_Y1}/\tau_{rad\_Y1} = 41.5\%$. To specify the $Y_2$-$Z_1$ branching ratio, further experiments are needed to distinguish emission between $Y_1$-$Z_1$ and $Y_2$-$Z_1$ and determine the phonon relaxation rate from $Y_2$-$Y_1$ as a function of temperature.

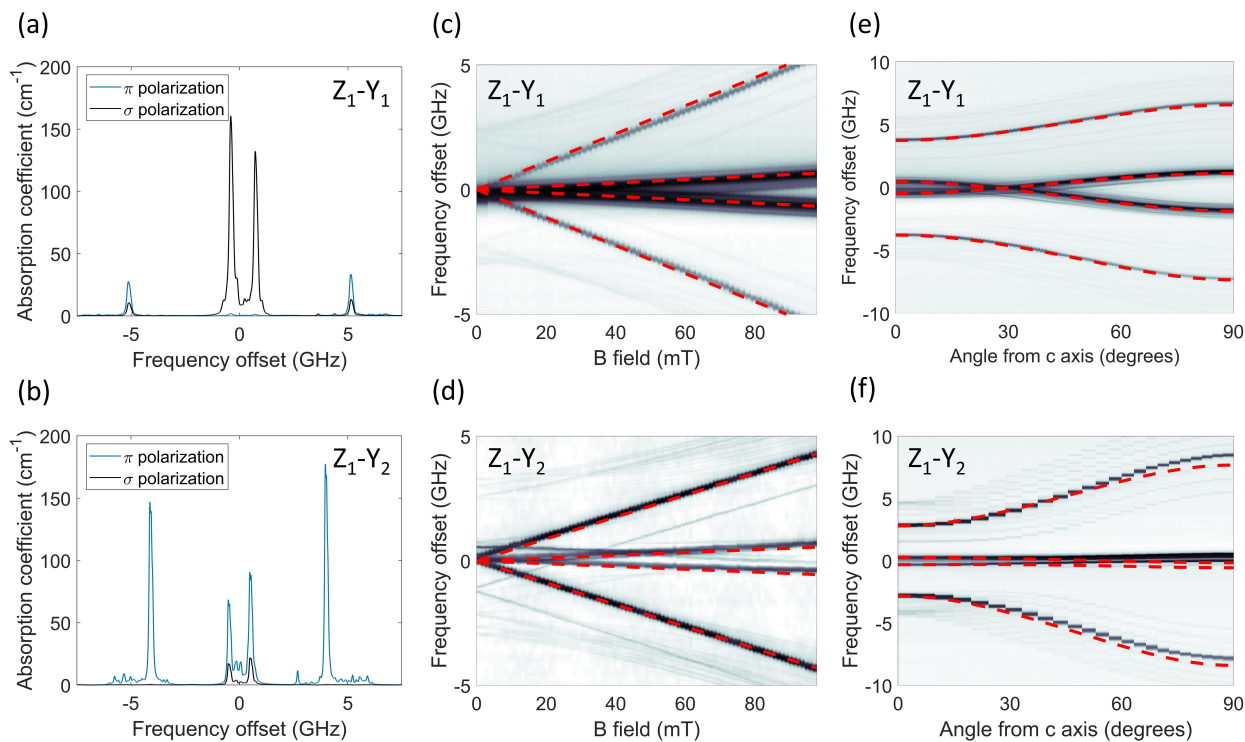

Fig. 4: The high-resolution optical transmission spectrum of 140 ppm $Er^{3+}$:$YVO_4$. (a) and (b) show the optical transmission spectrum of $Z_1$-$Y_1$ and $Z_1$-$Y_2$ respectively at an applied field of 90 mT at 1 K. The blue line labels $\pi$ ($\mathbf{E}_{ac}\|c$) polarization and the black line labels $\sigma$ ($\mathbf{E}_{ac}\|a$) polarization. (c) and (d) show the magnetic field ramp of $Z_1$-$Y_1$ and $Z_1$-$Y_2$ transmission spectra. The red dashed lines are the fitting from the effective spin Hamiltonian. (e) and (f) show the transmission spectra as a magnetic field of 75 mT is rotated from parallel to the crystal c-axis to parallel with crystal a-axis. The red dashed lines are the fitting from the effective spin Hamiltonian.

| Transitions | Wavelength (nm) | Optical Inhomogeneity (MHz) | Polarization | Dipole Type | $\int \alpha \, d\nu$ (GHz*cm$^{-1}$) | Oscillator Strength (1e-7) | Dipole Moment (1e-32 C m) | $1/\tau_{rad}$ (Hz) |
|---|---|---|---|---|---|---|---|---|
| $Z_1$-$Y_1$ | 1529.21 | 184 ± 10 | $\sigma$ | ED | 7.3 | 0.8 | 1 | 5.7 |
| | | | $\sigma$ | MD | 89.9 | 9.0 | 3.3 | 85 |
| | | | $\pi$ | MD | 18.0 | 2.0 | 1.6 | 13.9 |
| $Z_1$-$Y_2$ | 1528.78 | 163 ± 14 | $\sigma$ | ED | 10.7 | 1.2 | 1.2 | 8.3 |
| | | | $\pi$ | ED | 79.5 | 7.6 | 3.0 | 75.2 |
| | | | $\pi$ | MD | 45.5 | 5.1 | 2.5 | 35.3 |

Table 1: Optical properties of $Er^{3+}$:$YVO_4$ transitions including the transition wavelength, optical inhomogeneities, integrated absorption coefficients, oscillator strength, dipole moment and the radiative lifetime of $Z_1$-$Y_1$ and $Z_1$-$Y_2$ transitions.

| Levels | $|g_\parallel|$ | $|g_\perp|$ |
|---|---|---|
| $Z_1$ | 3.544[48] | 7.085[48] |
| $Y_1$ | 4.51 ± 0.02 | 4.57 ± 0.01 |
| $Y_2$ | 2.74 ± 0.01 | 6.74 ± 0.15 |

Table 2: The electronic Zeeman $g$ factors of $Er^{3+}$:$YVO_4$. The $g$ factors of the $Z_1$ level are from [48]. The $g$ factors of the $Y_1$ and $Y_2$ levels are from fitting to experimental data using the effective spin Hamiltonian.

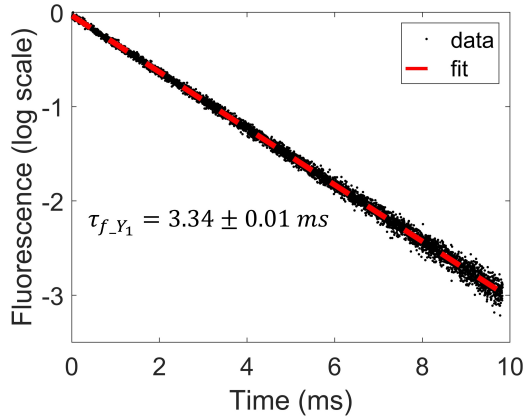

Fig. 5: $Y_1$ lifetime measurement via fluorescence decay at 1 K. Exponential fit is showed by the red dashed line, with decay constant $\tau_{f\_Y_1} = 3.34 \pm 0.01$ ms.

B. **Electron paramagnetic resonance measurement**

Here we study the spin inhomogeneity by measuring the EPR spectrum as a function of the applied magnetic field. We placed a 2 mm thick (10× thicker than the optical measurements) sample into the loop-gap microwave resonator ($B_{ac}$∥a). When the external magnetic field ($B_{dc}$∥c) induced splitting causes the $Z_1$ spin transition to be near the loop-gap resonance, the center frequency of the resonator is slightly shifted due to the dispersive coupling with the spin ensemble. The ensemble coupling is expected to be $\Omega = \mu_{21}\sqrt{\frac{\rho \Delta n(T) \eta \omega_o \mu_0}{2\hbar}}$, where $\mu_{21}$ is the magnetic dipole moment, $\rho$ is the number density of the ions, $\Delta n(T)$ is the population difference between the spin levels, $\omega_o$ is the center frequency of the resonator, $\eta$ is the magnetic energy fraction in the $Er^{3+}$:$YVO_4$ crystal, $\hbar$ is the reduced Planck constant, and $\mu_0$ is the vacuum permeability. From the simulation of the cavity, we predict $\eta \sim 2\%$, corresponding to an ensemble coupling $\Omega \approx 2.3$ MHz at 1 K. In the weak coupling regime, the dispersive shift is proportional to $\Omega^2/\Delta$ [51], where $\Delta$ is the spin inhomogeneity.

To increase our measurement sensitivity for dispersive shifts smaller than the linewidth of the microwave cavity (FWHM = 2.8 MHz), we implemented a FM microwave tone to monitor small changes in the microwave cavity resonance frequency [45]. To determine the spin inhomogeneity, we follow the analysis from [52], where a cavity with frequency $\omega_0$ couples to a distribution of N two-level systems with transition frequencies $\omega_k$ and damping rates $\gamma$ with strength $g_k$ in the weak excitation regime. The cavity transfer function is given by:

$$t(\omega) = \frac{\kappa/2i}{\omega - \omega_0 + i\kappa/2 - W(\omega)}, \quad W(\omega) = \Omega^2 \int_{-\infty}^{+\infty} \frac{\rho(\omega')}{\omega - \omega' + i\gamma/2} d\omega'$$

where $\kappa$ is the linewidth of the cavity, $\Omega^2 = \sum_k g_k^2$ is the total coupling between the cavity and the spins and $\rho(\omega) = \sum_k g_k^2 \delta(\omega - \omega_k)/\Omega^2$ is the spectral density distribution. For a Gaussian distribution, $\rho(\omega) = \frac{\sqrt{\ln 2}}{\Delta\sqrt{\pi}} e^{-(\omega^2 \ln 2)/\Delta^2}$, where $\Delta$ is the spin inhomogeneity.

The transmitted FM microwave field was measured on a microwave power meter at the modulation frequency $\omega_m$:

$$P_t = \text{DC part} + P_0\beta\{Re[\chi(\omega)]cos\omega_m t + Im[\chi(\omega)]sin\omega_m t\} + \text{terms oscillating at } 2\omega_m$$

Where $\beta$ is the modulation strength and $\chi(\omega) = t(\omega)t^*(\omega + \omega_m) - t^*(\omega)t(\omega - \omega_m)$. When $\omega_m \ll \kappa$ and $\kappa \ll \Delta$, $\chi(\omega)$ has the shape of the derivative of $|t(\omega)|^2$, which transforms the resonance peak to a zero-crossing point. We measured the beat-note signal on a spectral analyzer, taking advantage of the large dynamic range to maximize our frequency sensitivity. Using this technique, we measured kilohertz frequency shifts, which were ~1000 times smaller than the resonator linewidth.

As shown in Fig. 6(a), as the magnetic field was swept such that the spin frequency crossed the bare cavity frequency, we observed a dispersive response of the microwave resonator frequency. The strongest signal at 48mT is from even isotopes, where other coupling points at $\mathbf{B}_{dc}$ = 12, 41, 68 mT are attributed to $^{167}$Er isotope hyperfine structure. To understand the hyperfine coupling points, we simulate the $Z_1$ level hyperfine structure of $^{167}$Er using the parameters presented in [48]. The calculated hyperfine transitions are showed in Fig. 6(b), where the transition strength is evaluated by taking the inner product of the initial and final states mediated by a $\sigma_x \otimes I_N$ operator to take into account $\Delta m_s = \pm 1$ and $\Delta m_I = 0$ selection rules for $\mathbf{B}_{ac}\|a$. The red dashed line indicates the microwave resonator frequency. The black dashed line indicates the electronic spin frequency. There are three allowed hyperfine transitions crossing the 2.4 GHz loop-gap resonance at different bias magnetic fields, which is in agreement with the experimental data as indicated by the red dashed arrows.

Fig. 6(c) shows the detail of the scan around 48mT at 1 K, where the even isotope dispersive shift increased compared to Fig. 6(a) due to a bigger ground-state spin population difference at the lower temperature. We calculate the $Z_1$ electron spin inhomogeneity to be 58.4 MHz with an ensemble coupling of 3.1 MHz, assuming a Gaussian line shape of the spin inhomogeneity for fitting the data (see Fig. 6(c)). To make sure we were not limited by the power broadening and saturation effect on the spin coupling, we swept the input microwave power from -60 to -20 dBm, as shown in Fig. 6(d). We observe that the fitted spin inhomogeneity and ensemble coupling remain roughly constant as the input microwave power is smaller than -50dBm. However, the fitted 58.4 MHz spin inhomogeneity is an upper bound given that we expect spatial magnetic field inhomogeneity from our home-made magnet over the 3.5×2.5×2 mm sample volume inside the microwave resonator. We observed similar inhomogeneities at the $^{167}$Er hyperfine coupling points that scaled linearly with the effective g factor of each hyperfine transition, which suggests the inhomogeneity is limited by the magnetic field inhomogeneity.

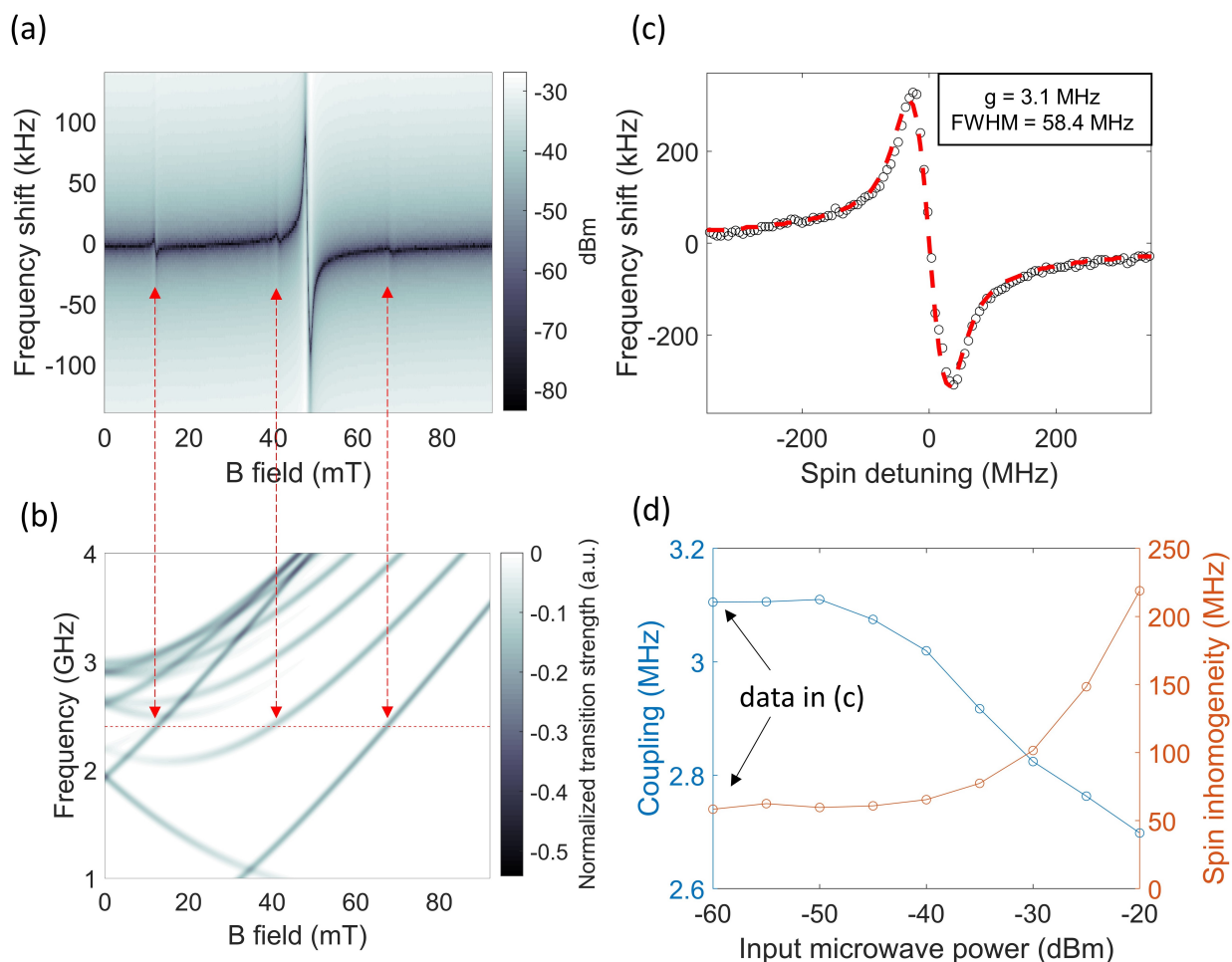

Fig. 6: The simulation and experimental results of the EPR measurements. (a) A full EPR spectrum taken at 4 K. The dispersive shift of the microwave resonator frequency indicates the coupling between the atomic ensemble and the resonator mode. The coupling at 48 mT is contributed from the even isotope of erbium with zero nuclear spin. The other three couplings located at 12 mT, 41 mT and 68 mT are from the $^{167}$Er with 7/2 nuclear spin. (b) A simulation

showing the allowed hyperfine transitions with $\Delta m_s = \pm 1$ and $\Delta m_I = 0$ selection rules under $B_{ac} \| a$ condition. The red dashed line indicates the microwave resonator center frequency. The black dashed line indicates the $Z_1$ spin transition frequency as the magnetic field varies. The red dashed arrows show the agreement of the $Z_1$ spin hyperfine structure between the experiment and the simulation. (c) A detailed scan of the spin ensemble-cavity coupling centered at 48 mT taken at a temperature of 1 K. The red dashed line is the fitting results using the model in [52]. (d) The power dependence of EPR results.

### C. <u>Microwave to optical transduction</u>

To generate upconverted optical photons, we input microwave signal and an optical pump into an $Er^{3+}$:YVO$_4$ crystal and measure the upconverted signal using heterodyne detection [45]. The key to this Raman scattering protocol is to have a 3-level system, where three atomic levels are linked via two allowed optical transitions and one allowed microwave transition. Using the $Er^{3+}$:YVO$_4$ $Z_1$-$Y_1$ or $Z_1$-$Y_2$ systems in a bias magnetic field, we have four possible optical transitions and two possible microwave transitions. From the $Er^{3+}$:YVO$_4$ selection rules described in the background section, both the $Z_1$-$Y_1$ transition with $\sigma$ polarized excitation or the $Z_1$-$Y_2$ transition with $\pi$ polarized excitation, are suitable for transduction. For either system, the ground (excited) state spins can be tuned onto resonance with the microwave resonator using the bias field to make use of a Λ(V) system.

To demonstrate M2O transduction, we used the $Z_1$-$Y_2$ system under $\pi$ polarization excitation with a 200 μm thick sample, driving with a fixed 2.4 GHz RF frequency at a fixed 1 dBm RF power input and 4 μW of optical pump power, as shown in Fig. 7. The four configurations that generate the transduction signal (Fig. 7(a), (b)) are apparent. Two of them are centered at B = 47 mT corresponding to the $Z_1$ spins involved in Λ systems. These two branches are offset by 1.8 GHz in optical frequency, which corresponds to the $Y_2$ spin splitting. The other two signals that appear at B = 61 mT are generated from the V systems involving the $Y_2$ spin, where the 3.1 GHz optical frequency offset is the $Z_1$ spin splitting.

In $Er^{3+}$:YVO$_4$, the $Z_1$-$Y_2$ system is more attractive than $Z_1$-$Y_1$ because the two involved optical transitions have stronger transition strength (Table 1) and the transduction efficiency scales as the product square of the individual dipole moments in the limit of restricted optical pump power. As a result, the $Z_1$-$Y_2$ system should be more efficient by a factor of 6. By optimizing the applied magnetic field strength and optical frequency, we measured the transduction signal using $Z_1$-$Y_1$ and $Z_1$-$Y_2$ systems and observed that the relative signal strength is consistent with the prediction, as shown in Fig. 7(c). The trend in efficiency versus input microwave frequency follows the loop-gap resonator line-shape that determines the current transduction bandwidth. The inset in Fig. 7(c) shows the optical pump power ($P_o$) dependence of $Z_1$-$Y_2$ system transduction efficiency. Up to 300 μW input power, limited by the setup, the transduction efficiency increases linearly. Finally, the highest measured efficiency of this initial demonstration of transduction process is $1.3 \times 10^{-12}$ under 1 dBm microwave and 300 μW optical pump power. To further increase the efficiency, a better microwave resonator with higher Q and incorporating with an optical cavity are needed as discussed in [18,20].

To compare with other rare-earth materials for M2O transduction application, we refer to the reference [14] where a cavity enhanced M2O transduction experiment of a 10 ppm $Er^{3+}$:$Y_2SiO_5$ and all relevant spectroscopic parameters are presented. To evaluate the M2O transduction process with a 3-level system, we compare the material parameter $\zeta = \left(\frac{d_{31}d_{32}\mu_{21}\rho}{\Delta_o \Delta_\mu}\right)^2$ which scales as the efficiency (when $\eta < 1\%$). $\zeta$ for $Er^{3+}$:$YYO_4$ $Z_1$-$Y_2$ system is 800× larger due to its relatively narrow optical and spin inhomogeneities at higher concentration and also the stronger dipole moment. However, other spectroscopic research [28] shows that $Er^{3+}$:$Y_2SiO_5$ can have relatively narrow optical inhomogeneity at a higher concentration (510 MHz at 200 ppm compared to 170 MHz at 10ppm). Assuming the spin inhomogeneity unchanged, the $\zeta_{Er^{3+}:YVO_4}/\zeta_{Er^{3+}:Y_2SiO_5}$ ratio will reduce to 18 from 800. Therefore, both $Er^{3+}$:$Y_2SiO_5$ and $Er^{3+}$:$YYO_4$ materials can potentially be improved at higher concentration if they still possess relatively narrow optical and spin inhomogeneities. Finally, the 58.4 MHz $Er^{3+}$:$YYO_4$ spin inhomogeneity measured here is limited by the spatial magnetic field inhomogeneity, which makes the above $\zeta_{Er^{3+}:YVO_4}$ analysis as a lower bound. In conclusion, the relatively narrow ensemble inhomogeneity and strong dipole moment make $Er^{3+}$:$YYO_4$ promising for the transduction application.

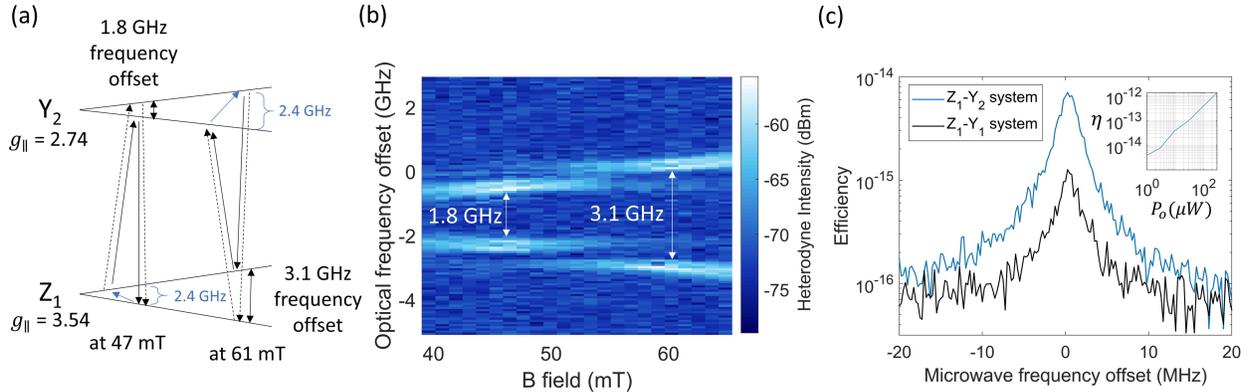

Fig. 7: The microwave to optical transduction experiment using Raman heterodyne detection, with the $Z_1$-$Y_2$ and $Z_1$-$Y_1$ systems. (a) The $Z_1$-$Y_2$ system used in (b). For example, at 47 mT, the $Z_1$ spins are matched to be on resonance with the 2.4 GHz microwave resonator. Two $\lambda$ systems are usable for transduction with a 1.8 GHz optical frequency offset. (b) The Raman heterodyne spectroscopy using the $Z_1$-$Y_2$ system. Two signal branches at 47 mT correspond to two $\Lambda$ systems and the other two branches at 61 mT correspond to two V systems. (c) Transduction efficiency versus input microwave frequency using $Z_1$-$Y_1$ and $Z_1$-$Y_2$ systems. The current bandwidth is limited by the loop-gap microwave resonator linewidth. The inset figure shows the efficiency ($\eta$) dependence upon optical pump power ($P_o$) using the $Z_1$-$Y_2$ system.

## V: Conclusion:

In this work, we present the experimental characterization of $Er^{3+}$:$YVO_4$ including the electronic Zeeman $g$ tensors of the optically excited states $Y_1$ and $Y_2$, the optical properties of the telecommunication wavelength transitions $Z_1$-$Y_1$ and $Z_1$-$Y_2$, the spin inhomogeneity of the $Z_1$ spin transition and an initial demonstration of microwave-to-optical transduction using $Er^{3+}$:$YVO_4$. The optical inhomogeneity of a 140 ppm sample is narrow (~163 MHz) and the total dipole moments are relatively strong compared to other erbium-doped materials ($3.5 \times 10^{-32}$ Cm for electric-dipole transitions and $3.5 \times 10^{-32}$ Cm for magnetic-dipole transitions in $Z_1$-$Y_2$ system). We also measure a spin inhomogeneity of 58.4 MHz as an upper bound. Therefore, the highly absorptive optical transitions and the narrow inhomogeneities make $Er^{3+}$:$YVO_4$ a promising material for microwave to optical transduction applications. Next steps towards more efficient M2O transduction will involve using $Er^{3+}$:$YVO_4$ in the cavity enhanced regime in both optical and spin domains.

## **Acknowledgements:**
This work was supported by Army Research Office (ARO/LPS) (CQTS) grant number W911NF1810011, NSF QII TAQS 1936350, Office of Naval Research Award No. N00014-19-1-2182, Air Force Office of Scientific Research grant numbers FA9550-18-1-0374 and FA9550-21-1-0055, Northrop Grumman and Weston Havens Foundation. J.G.B. acknowledges the support of the American Australian Association's Northrop Grumman Fellowship. I.C. and J.R. acknowledge support from the Natural Sciences and Engineering Research Council of Canada (Grant Nos. PGSD2-502755-2017 and PGSD3-502844-2017). We would also like to acknowledge E. Miyazono for the initial measurements of $Er^{3+}$:$YVO_4$ and M. Shaw for help with superconducting photon detectors.

## **Reference:**
[1] Arute, Frank, et al. "Quantum supremacy using a programmable superconducting processor." Nature 574.7779 (2019): 505-510.
[2] Bruzewicz, Colin D., et al. "Trapped-ion quantum computing: Progress and challenges." Applied Physics Reviews 6.2 (2019): 021314.
[3] Yin, Juan, et al. "Satellite-based entanglement distribution over 1200 kilometers." Science 356.6343 (2017): 1140-1144.
[4] Pompili, Matteo, et al. "Realization of a multi-node quantum network of remote solid-state qubits." arXiv preprint arXiv:2102.04471 (2021).
[5] Heshami, Khabat, et al. "Quantum memories: emerging applications and recent advances." Journal of modern optics 63.20 (2016): 2005-2028.
[6] Yu, Yong, et al. "Entanglement of two quantum memories via fibres over dozens of kilometres." Nature 578.7794 (2020): 240-245.
[7] Kumar, Sourabh, Nikolai Lauk, and Christoph Simon. "Towards long-distance quantum networks with superconducting processors and optical links." Quantum Science and Technology 4.4 (2019): 045003.
[8] Lauk, Nikolai, et al. "Perspectives on quantum transduction." Quantum Science and Technology 5.2 (2020): 020501.
[9] Lambert, Nicholas J., et al. "Coherent conversion between microwave and optical photons—an overview of physical implementations." Advanced Quantum Technologies 3.1 (2020): 1900077.
[10] Higginbotham, Andrew P., et al. "Harnessing electro-optic correlations in an efficient mechanical converter." Nature Physics 14.10 (2018): 1038-1042.
[11] Mirhosseini, Mohammad, et al. "Superconducting qubit to optical photon transduction." Nature 588.7839 (2020): 599-603.

[12] Rueda, Alfredo, et al. "Efficient microwave to optical photon conversion: an electro-optical realization." Optica 3.6 (2016): 597-604.
[13] Fan, Linran, et al. "Superconducting cavity electro-optics: a platform for coherent photon conversion between superconducting and photonic circuits." Science advances 4.8 (2018): eaar4994.
[14] Fernandez-Gonzalvo, Xavier, et al. "Cavity-enhanced Raman heterodyne spectroscopy in Er 3+: Y 2 SiO 5 for microwave to optical signal conversion." Physical Review A 100.3 (2019): 033807.
[15] Han, Jingshan, et al. "Coherent microwave-to-optical conversion via six-wave mixing in Rydberg atoms." Physical review letters 120.9 (2018): 093201.
[16] Hisatomi, Ryusuke, et al. "Bidirectional conversion between microwave and light via ferromagnetic magnons." Physical Review B 93.17 (2016): 174427.
[17] Das, Sumanta, et al. "Interfacing superconducting qubits and single optical photons using molecules in waveguides." Physical review letters 118.14 (2017): 140501.
[18] Williamson, Lewis A., Yu-Hui Chen, and Jevon J. Longdell. "Magneto-optic modulator with unit quantum efficiency." Physical review letters 113.20 (2014): 203601.
[19] O'Brien, Christopher, et al. "Interfacing superconducting qubits and telecom photons via a rare-earth-doped crystal." Physical review letters 113.6 (2014): 063603.
[20] Bartholomew, John G., et al. "On-chip coherent microwave-to-optical transduction mediated by ytterbium in yvo 4." Nature communications 11.1 (2020): 1-6.
[21] Everts, Jonathan R., et al. "Microwave to optical photon conversion via fully concentrated rare-earth-ion crystals." Physical Review A 99.6 (2019): 063830
[22] Thiel, C. W., Thomas Böttger, and R. L. Cone. "Rare-earth-doped materials for applications in quantum information storage and signal processing." Journal of luminescence 131.3 (2011): 353-361.
[23] Kindem, Jonathan M., et al. "Characterization of Yb 3+ 171: YVO 4 for photonic quantum technologies." Physical Review B 98.2 (2018): 024404.
[24] Zhong, Manjin, et al. "Optically addressable nuclear spins in a solid with a six-hour coherence time." Nature 517.7533 (2015): 177-180.
[25] Liu, Guokui, and Bernard Jacquier, eds. Spectroscopic properties of rare earths in optical materials. Vol. 83. Springer Science & Business Media, 2006.
[26] Rakonjac, Jelena V., et al. "Long spin coherence times in the ground state and in an optically excited state of Er 3+ 167: Y 2 SiO 5 at zero magnetic field." Physical Review B 101.18 (2020): 184430.
[27] Sun, Yongchen, et al. "Magnetic g tensors for the I 15∕2 4 and I 13∕2 4 states of Er 3+: Y 2 Si O 5." Physical Review B 77.8 (2008): 085124.
[28] Böttger, Thomas, et al. "Spectroscopy and dynamics of Er 3+: Y 2 Si O 5 at 1.5 μ m." Physical Review B 74.7 (2006): 075107.
[29] Rančić, Miloš, et al. "Coherence time of over a second in a telecom-compatible quantum memory storage material." Nature Physics 14.1 (2018): 50-54.
[30] Popova, M. N., et al. "Crystal field and hyperfine structure of Er 3+ 167 in YP O 4: Er single crystals: High-resolution optical and EPR spectroscopy." Physical Review B 99.23 (2019): 235151.
[31] Fukumori, Rikuto, et al. "Subkilohertz optical homogeneous linewidth and dephasing mechanisms in Er 3+: Y 2 O 3 ceramics." Physical Review B 101.21 (2020): 214202.

[32] Marino, Robert, et al. "Energy level structure and optical dephasing under magnetic field in Er3+: LiYF4 at 1.5 μm." Journal of Luminescence 169 (2016): 478-482.
[33] Jiang, Xiaodong, et al. "Rare earth-implanted lithium niobate: Properties and on-chip integration." Applied Physics Letters 115.7 (2019): 071104.
[34] Phenicie, Christopher M., et al. "Narrow optical line widths in erbium implanted in TiO2." Nano letters 19.12 (2019): 8928-8933.
[35] Capobianco, J. A., et al. "Optical spectroscopy, fluorescence dynamics and crystal-field analysis of Er3+ in YVO4." Chemical physics 214.2-3 (1997): 329-340.
[36] Ter-Gabrielyan, Nikolay, et al. "Spectroscopic and laser properties of resonantly (in-band) pumped Er: YVO 4 and Er: GdVO 4 crystals: a comparative study." Optical Materials Express 2.8 (2012): 1040-1049.
[37] Le Boulanger, Ph, et al. "Excited-state absorption spectroscopy of Er 3+-doped Y 3 Al 5 O 12, YVO 4, and phosphate glass." Physical Review B 60.16 (1999): 11380.
[38] Golab, Stanislaw, et al. "Relaxation dynamics of excited states of Er3+ in YVO4 single crystals." International Conference on Solid State Crystals 2000: Growth, Characterization, and Applications of Single Crystals. Vol. 4412. International Society for Optics and Photonics, 2001.
[39] Li, Pei-Yun, et al. "Optical spectroscopy and coherent dynamics of 167Er3+: YVO4 at 1.5 μm below 1 K." Journal of Luminescence (2020): 117344.
[40] Chakoumakos, Bryan C., Marvin M. Abraham, and Lynn A. Boatner. "Crystal structure refinements of zircon-type MVO4 (M= Sc, Y, Ce, Pr, Nd, Tb, Ho, Er, Tm, Yb, Lu)." Journal of Solid State Chemistry 109.1 (1994): 197-202.
[41] Abragam, Anatole, and Brebis Bleaney. Electron paramagnetic resonance of transition ions. OUP Oxford, 2012.
[42] Koster, George F., et al. Properties of the thirty-two point groups. Vol. 24. MIT press, 1963.
[43] Luo, Zundu, and Yidong Huang. "Physics of Solid-State Laser Materials." Springer Series in Materials Science, Springer Singapore, 2020.
[44] Hufner, Stefan, ed. Optical spectra of transparent rare earth compounds. Elsevier, 2012.
[45] Fernandez-Gonzalvo, Xavier, et al. "Coherent frequency up-conversion of microwaves to the optical telecommunications band in an Er: YSO crystal." Physical Review A 92.6 (2015): 062313.
[46] Shi, H-S., G. Zhang, and H-Y. Shen. "Measurement of principal refractive indices and the thermal refractive index coefficients of yttrium vanadate." Journal of Synthetic Crystals 30.1 (2001): 85-88.
[47] Dodson, Christopher M., and Rashid Zia. "Magnetic dipole and electric quadrupole transitions in the trivalent lanthanide series: Calculated emission rates and oscillator strengths." Physical Review B 86.12 (2012): 125102.
[48] Ranon, U. "Paramagnetic resonance of Nd3+, Dy3+, Er3+ and Yb3+ in YVO4." Physics Letters A 28.3 (1968): 228-229.
[49] Veissier, Lucile, et al. "Quadratic Zeeman effect and spin-lattice relaxation of Tm 3+: YAG at high magnetic fields." Physical Review B 94.20 (2016): 205133.
[50] Sumida, D. S., and T. Y. Fan. "Effect of radiation trapping on fluorescence lifetime and emission cross section measurements in solid-state laser media." Optics Letters 19.17 (1994): 1343-1345.
[51] Dold, Gavin, et al. "High-cooperativity coupling of a rare-earth spin ensemble to a superconducting resonator using yttrium orthosilicate as a substrate." Physical Review Applied 11.5 (2019): 054082.

[52] Diniz, I., et al. "Strongly coupling a cavity to inhomogeneous ensembles of emitters: Potential for long-lived solid-state quantum memories." Physical Review A 84.6 (2011): 063810.